\documentclass[sn-mathphys,Numbered]{sn-jnl}% Math and Physical Sciences Reference Style
\usepackage{graphicx}%
\usepackage{multirow}%
\usepackage{amsmath,amssymb,amsfonts}%
\usepackage{amsthm}%
\usepackage{mathrsfs}%
\usepackage[title]{appendix}%
\usepackage{xcolor}%
\usepackage{textcomp}%
\usepackage{manyfoot}%
\usepackage{booktabs}%
\usepackage{algorithm}%
\usepackage{algorithmicx}%
\usepackage{algpseudocode}%
\usepackage{listings}%
\theoremstyle{thmstyleone}%
\theoremstyle{thmstyletwo}%

\theoremstyle{thmstylethree}%

\begin{document}

\title[Article Title]{Quantum Prime Factorization: A Novel Approach Based on Fermat's Method
}


\author*[1]{\fnm{Julien} \sur{Mellaerts}}\email{julien.mellaerts@eviden.com}

\affil*[1]{\orgname{Eviden}, \orgaddress{\city{Les Clayes-sous-Bois}, \country{France}}}


\abstract{
In this paper, we introduce a novel quantum algorithm for the factorization of composite odd numbers. This work makes two significant contributions. First, we present a new improvement to the classical Fermat's method, fourfold reducing the computational complexity of factoring. Second, we reformulate Fermat's factorization method as an optimization problem suitable for Quantum Annealers which allowed us to factorize 8,689,739, the biggest number ever factorized using a quantum device to our knowledge.
}

\maketitle

\section{Introduction}\label{sec1}

The integer factorization problem, which involves finding the prime factors of a composite number, is a cornerstone of modern cryptography. This problem is particularly significant in the context of RSA encryption\citep{10.1145/359340.359342}, a widely used public-key cryptosystem that relies on the difficulty of factoring large integers into their prime components. In RSA, the security is predicated on the assumption that, given a large composite number $N$, which is the product of two large prime numbers $p$ and $q$, it is computationally difficult to determine $p$ and $q$ using classical computing methods. This assumption underpins the security of countless secure communications and data transmissions worldwide. 

\bigskip\noindent
The advent of quantum computing has introduced new paradigms for solving complex computational problems, including integer factorization. Among the most notable quantum algorithms is Shor's algorithm\citep{doi:10.1137/S0097539795293172}, developed by Peter Shor in 1994. Shor's algorithm leverages the principles of quantum mechanics to factor large integers in polynomial time, offering an exponential speedup over the best-known classical algorithms such as the General Number Field Sieve\citep{10.1145/100216.100295}. This capability poses a significant threat to RSA encryption, as it could potentially render current cryptographic methods obsolete if large-scale, fault-tolerant quantum computers become a reality.

\bigskip\noindent
In addition to Shor's algorithm, other quantum approaches have been explored for integer factorization, including those based on quantum annealing\citep{article1, article2}. Quantum annealing is a metaheuristic for finding the global minimum of an optimization problem and has been implemented in devices like the D-Wave quantum computer.

\bigskip\noindent
Classically, Fermat's factorization method is a well-known technique for factoring composite odd integers. It is based on the representation of an odd integer $N$ as a difference of two squares. While Fermat's method is elegant and straightforward, it is not efficient for very large numbers, which limits its practical application in modern cryptography.

\bigskip\noindent
This paper first review classical Fermat's method for factorization and propose an improvement fourfold reducing the number of iterations, secondly, it reformulates it as a combinatorial optimization problem suitable for quantum annealers. Instead of directly looking for $p$ and $q$ such that $N=p\cdot q$ like in previous work, where the multiplication increases the number of variables because of carries encoding and degree reduction, the Fermat's method reduces the problem as a difference of two squares, avoiding ancillary variables. By leveraging the capabilities of quantum annealing, this idea aims to shift the complexity from iteration count of the classical Fermat's method to variable count to enhance the efficiency and scalability of Fermat's method.

\section{Fermat’s Method}\label{sec2}

\subsection{Initial Fermat's Formulation}\label{subsec21}
Fermat’s factorization method is based on the fact that any composite odd integer $N$ can be represented as the difference of two squares:
$$
N = x^2 - y^2 = (x-y)(x+y)
$$
Where $N=p\cdot q$ for two primes $p$ and $q$, and
$$
x=\frac{p+q}{2},\quad y=\frac{p-q}{2}
$$

\bigskip\noindent
The purpose of the algorithm is to search for integers $x$ and $y$ such that $x^2-N = y^2$ is a perfect square, it proceeds by starting with $x=\lceil\sqrt{N}\rceil$ and incrementing $x$ by one until $x^2-N$ is a perfect square. Once $x$ and $y$ are known, $p$ and $q$ can be computed as 
$$
p=x+y,\quad q=x-y
$$

\bigskip\noindent
From Fermat's method, we know that the lower bound of $x$ is $x_{min}=\lceil\sqrt{N}\rceil$, and the upper bound of $x$ is $x_{max}=\frac{N+\delta_{min}^2}{2\delta_{min}}$ where $\delta_{min}$ is the minimal prime factor. Theoretically, we have $\delta_{min}=3$, as 3 is the smallest possible prime, thus the theoretical upper bound of $x$ is $\frac{N+9}{6}$. 

\noindent
Considering the case of RSA prime factorization, it does not make sense to suppose the smallest possible prime to be 3, as recommendations are, given $N$ a composite odd number of bitlength $n$, $p$ and $q$ should have $n/2$ bitlength. We can then consider without loss of generality that $\delta_{min}=2^{n/2-1}+1$ (as it must be odd). The lower bound of $y$ is $y_{min}=0$, the upper bound of $y$, $y_{max}$ can be deducted from $x_{max}$: $y_{max}=\sqrt{x_{max}^2-N}$

\subsection{Fermat's Method Improvements}\label{subsec22}

Since its initial formulation by Fermat, in a letter to Mersenne around 1643, several improvements have been made, based on different concepts such that replacing the high-cost operation (the perfect square in Fermat’s method), with a low-cost operation, or reducing the search space to find the solution.

\bigskip\noindent
The first most significant improvement used in this work is from Xiang\citep{xiang}, where he notices that we can deduce the parity of $x^2$ depending on the form of the composite odd integer $N$. Indeed, if $N$ is odd we have:
\begin{equation}
N \equiv \pm 1 \mod 4
\end{equation}

\noindent
And for any integer $n$ we know that:
\begin{equation}
\forall n \in \mathbb{N} \left\{
    \begin{array}{ll}
        even \Leftrightarrow n^2 \equiv 0 \mod 4 \\
        odd  \Leftrightarrow n^2 \equiv 1 \mod 4 
    \end{array}
\right.
\end{equation}
\noindent
Let $N$ be an odd composite, from Fermat's method we have: $ \exists x, y \in \mathbb{N} : N = x^2 - y^2 $.

\bigskip\noindent
Using (1) and by exhaustion:

\begin{unenumerate}[1.]
\item If $N \equiv 1 \mod 4$, $ x^2 - y^2 \equiv 1 \mod 4$, and by (2) $\left\{
    \begin{array}{ll}
        \text{x is odd} \\
        \text{y is even}
    \end{array}
    \right.$
\item If $N \equiv -1 \mod 4$, $ x^2 - y^2 \equiv -1 \mod 4$, and by (2) $\left\{
    \begin{array}{ll}
        \text{x is even} \\
        \text{y is odd}
    \end{array}
    \right.$
\end{unenumerate}

\noindent
Therefore, instead of increasing by one at each iteration, we can increase by two to look for either odd or even perfect square depending on the residue of $N \mod 4$, halving the number of iterations.

\bigskip\noindent
An other improvement used in this work is from Somsuk et al.\citep{inbook}, considering that for any prime number $p>3$, $p \equiv \pm 1 \mod 6$. Let $N=p \cdot q$, where $p, q>3$ both primes, we have:
\begin{unenumerate}[1.]
\item If $p \equiv \pm 1 \mod 6$ and $q \equiv \pm 1 \mod 6$, then $N=p \cdot q \equiv 1 \mod 6$, and $p+q \equiv \pm 2 \mod 6$
\item If $p \equiv \pm 1 \mod 6$ and $q \equiv \mp 1 \mod 6$, then $N=p \cdot q \equiv -1 \mod 6$, and $p+q \equiv 0 \mod 6$
\end{unenumerate}
Therefore, the incrementing step of $x$ can be conditioned by the residue of $N \ mod 6$. 

\section{Proposed Fermat’s Method Improvement}\label{sec3}

We have previously seen that, given $N$ an odd composite integer we have $N \equiv \pm 1 \mod 4$. The proposed improvement considers $k \mod 2$ as conditions to reduce the number of iterations, where:
$$
k \in \mathbb{N} \left\{
    \begin{array}{ll}
        k=\frac{N+1}{4} \text{ if } N \equiv -1 \mod 4 \\
        k=\frac{N-1}{4} \text{ if } N \equiv 1 \mod 4 
    \end{array}
\right.
$$

\addcontentsline{toc}{section}{nmin1}
\subsubsection*{Case of $N \equiv -1 \mod 4$}
for any integer $n$ we know that:
\begin{equation}
\forall n \in \mathbb{N} \left\{
    \begin{array}{ll}
        \text{even} \Leftrightarrow n^2 \mod 8 \in \{0, 4\}\\
        \text{odd}  \Leftrightarrow n^2 \mod 8 \in \{1\}
    \end{array}
\right.
\end{equation}
If $N \equiv -1 \mod 4$ we have:
\begin{equation}
N \mod 8 \in \{3, 7 \}
\end{equation}

\bigskip\noindent
From Fermat's method, we know that $ \exists x, y \in \mathbb{N} : N = x^2 - y^2 $, so we can write:
$$
(x^2 \mod 8) - (y^2 \mod 8) = (7-4\cdot(k \mod 2)) \mod 8
$$
Since $x^2$ is even and $y^2$ is odd, from (3) we know that $y^2 \equiv 1 \mod 8$:
$$
(x^2 \mod 8) - (1 \mod 8) = (7-4\cdot(k \mod 2)) \mod 8
$$
$$
(x^2 \mod 8) = ((7-4\cdot(k \mod 2)) \mod 8) + (1 \mod 8)
$$

\bigskip\noindent
We can see that we can consider only even perfect squares that comply with the residue of $N \mod 8$ conditioned by $k \mod 2$, which fourfold reduces the overall number of iterations.

\addcontentsline{toc}{section}{nplus1}
\subsubsection*{Case of $N \equiv 1 \mod 4$}
for any integer $n$ we know that:
\begin{equation}
\forall n \in \mathbb{N} \left\{
    \begin{array}{ll}
        even \Leftrightarrow n^2 \mod 16 \in \{0, 4\}\\
        odd  \Leftrightarrow n^2 \mod 16 \in \{1, 9\}
    \end{array}
\right.
\end{equation}
If $N \equiv 1 \mod 4$ we have:
\begin{equation}
N \mod 16 \in \{1, 5, 9, 13 \}
\end{equation}
\bigskip\noindent
From Fermat's method, we know that $ \exists x, y \in \mathbb{N} : N = x^2 - y^2 $, so we can write:
$$
(x^2 \mod 16) - (y^2 \mod 16) = N \mod 16
$$
Since $y^2$ is even, from (5) we know that $y^2 \mod 16 \in \{0, 4\}$:
$$
(x^2 \mod 16) - (0+4\cdot (k \mod 2) \mod 16) = N \mod 16
$$
$$
(x^2 \mod 16) = (N \mod 16) + (0+4\cdot (k \mod 2) \mod 16)
$$

\bigskip\noindent
We can see that we can consider only odd perfect squares that comply with the residue of $N \mod 16$ conditioned by $k \mod 2$, which also fourfold reduces the overall number of iterations.

\section{Quantum Reformulation of Fermat's Algorithm}\label{sec4}
This section describes the reformulation of Fermat's algorithm as a combinatorial optimization problem. Two different problem encoding approaches are proposed and discussed. 

\bigskip\noindent
Combinatorial optimization problems involve finding an optimal solution from a finite set of possible solutions. These problems are characterized by a discrete set of variables and constraints, and the goal is to find the best combination of these variables that satisfies the constraints and optimizes (either minimizes or maximizes) a given objective function.

\bigskip\noindent
Many combinatorial optimization problems can be mapped to the Ising model, a mathematical model of ferromagnetism in statistical mechanics. The Ising model consists of spins that can be in one of two states (+1 or -1), and the goal is to find the spin configuration that minimizes the energy of the system. Quantum annealers are naturally suited to solve Ising model problems, making them well-suited for combinatorial optimization. The Hamiltonian (energy function) of the Ising model is given by:
$$
H_{ising}=\sum\limits_{i}h_i\sigma_i + \sum\limits_{i<j}J_{i,j}\sigma_i\sigma_j
$$
Where $h_i$ is the external magnetic field on spin $i$ and $J_{ij}$ is the interaction strength between spins $i$ and $j$.

\bigskip\noindent
Equivalently, the Quadratic Unconstrained Binary Optimization (QUBO) model can be used to formulate combinatorial optimization problems. The Ising model and the QUBO model can be mapped to each other through a simple transformation of variables. Specifically, we can convert an Ising model to a QUBO model by using the following substitution:
$$
x_i=\frac{1+\sigma_i}{2}
$$
This transformation maps the spin variables $\sigma_i$ (which take values $+1$ or $-1$) to binary variables $x_i$ (which take values $0$ or $1$). 

\bigskip\noindent
The Hamiltonian of the QUBO model is given by:
$$
H_{QUBO}=\sum\limits_{i}Q_ix_i + \sum\limits_{i< j}Q_{i,j}x_ix_j
$$
Where $x_i$ are binary variables and $Q\in\mathbb{R}^{n\times n}$ is a square matrix. 

\bigskip\noindent
In this work, we will reformulate the Fermat's algorihtm as a QUBO model, which can be solved by a quantum annealer, a quantum-inspired or classical solver.

\addcontentsline{toc}{section}{The Objective Function}
\subsubsection*{The Objective Function}
From Fermat's method we have:
$$
x^2-y^2=N
$$
Thus
$$
x^2-y^2-N=0
$$
Which can be expressed as a cost function to minimize:
$$
min f(x,y)=(x^2-y^2-N)^2
$$

\medskip
\subsection{Perfect Square as Sum of Odds Based Approach}\label{subsec41}
The purpose of this approach is to use the relationship between perfect squares and the sum of odd numbers. Indeed, we have:
$$
r^2=\sum\limits_{i=1}^{r}2i-1
$$
e.g:
\begin{unenumerate}[3.]
\item $1^2=1$
\item $2^2=1+3$
\item $3^2=1+3+5$
\item $4^2=1+3+5+7$
\item $5^2=1+3+5+7+9$
\end{unenumerate}
By using this formalism, we can encode perfect squares $x^2$ and $y^2$ as two single sums.

\bigskip\noindent
To encode $x^2$, let us introduce $x_{lo}=\lceil\sqrt{N}\rceil^2$, $x_{up}=\lfloor\ \frac{N-\delta_{min}^2}{2\delta_{min}} \rfloor$, where $\delta_{min}=2^{(log_2(\sqrt{N})-1)}+1$, and $x'=\left\{ 2(x_{lo}+1)-1, 2(x_{lo}+2)-1,..., 2(x_{lo}+x_{up})-1\right\}$. Thus, any sum $x_{lo}+x'(0)$, $x_{lo}+x'(0)+x'(1)$... is a perfect square.

\bigskip\noindent
To encode $y^2$, let us introduce $y_{up}=\lfloor\ \sqrt{x_{up}^2-N} \rfloor$, $y'=\left\{ 2(1)-1, 2(2)-1,..., 2(y_{up})-1\right\}$. Thus, any sum $y'(0)$, $y'(0)+y'(1)$... is a perfect square.

\bigskip\noindent
The cost function we need to minimize is now:
$$
min f(x',y')=(x_{lo}+x'-y'-N)^2
$$
\bigskip\noindent
The QUBO matrix can be filled with terms $x'^2+y'^2-2x'y'+2x'N-2y'N$. 

\medskip\noindent
To ensure only consecutive odds to be selected, we need to add following constraints to the QUBO formulation:
$$
x_i'\geq x_{i+1}', \quad y_i'\geq y_{i+1}'
$$

\bigskip\noindent
These constraints are encoded in the QUBO matrix by adding the penalty:
$$
P(x_{i+1} - x_{i+1}x_{i}), \quad P(y_{i+1} - y_{i+1}y_{i})
$$

\bigskip\noindent
We can notice that this approach is made of linear and quadratic terms only, avoiding higher degrees.

\addcontentsline{toc}{section}{Complexity analysis}
\subsubsection*{Complexity analysis}

The space complexity of this approach is closely related to the number of iterations of the Fermat's method given by $x_{max}-\lceil N \rceil$, which in the worst case and in the context of RSA is $\mathcal{O}{(2^{\frac{n}{2}}-1-\sqrt{N})}$, where $n$ is the bitlength of $N$, corresponding to the space complexity of $x'$. For $y'$, the space complexity is $\mathcal{O}{(2^{\frac{n}{2}-2}-1)}$. The overall space complexity for this approach is $\mathcal{O}{(5\cdot2^{\frac{n}{2}-2}-\sqrt{N})}$.

\bigskip\noindent
This complexity does not take into account previously mentioned Fermat's method improvements. Even if it were the case, the number of variables required is too huge to be worth running on a real QPU. Nonetheless, this method has factorized a 32 bits odd composite using a classical solver on a laptop. 

\subsection{Perfect Square Pattern Based Approach}\label{subsec42}
This second encoding approach is based on the fact that the binary representation of perfect squares has some patterns. The binary representation of the square of an even number ends by '$00$', and the binary representation of the square of an odd number ends by '$01$'. More precisely, the binary representation of the square of an even number ending by '$1+k0$' (here $+$ refers to string concatenation and $k$ is the number of repetitions) ends by '$1+2k0$'; and the binary representation of the square of an odd number ending by '$1+k0+1$' ends by '$1+(k+1)×0+1$'. This property can be extended with the proposed improvement in section 3.

\bigskip\noindent
That said, we can define variables of our problem. we can define binary ending patterns of bitlength $l_{p,x}$ and $l_{p,y}$ for both binary representations of $x^2$ and $y^2$ respectively, $l_{x}=log(N)-l_{p,x}$ and $l_{y}=log(N)-l_{p,y}$ are bitlengths of $x^2$ and $y^2$, thus we can define $x'=(2^{l_{x}-1}x_{l_{x}-1}, 2^{l_{x}-2}x_{l_{x}-2},...,2^{l_{p,x}-1}x_{l_{p,x}-1})$, $y'=(2^{l_{y}-1}y_{l_{y}-1}, 2^{l_{y}-2}y_{l_{y}-2},...,2^{l_{p,y}-1}y_{l_{p,y}-1})$, where $x_i,y_i=\{0, 1\}$. Here, $x'$ and $y'$ sets denote binary encoding of $x^2$ and $y^2$ as we fix a perfect square ending pattern.

\noindent
The cost function we need to minimize is now:
$$
min f(x',y')=(x'-y'-N)^2
$$

\bigskip\noindent
The QUBO matrix can be filled with terms $x'^2+y'^2-2x'y'-2x'N+2y'N$. 

\bigskip\noindent
We can notice that this encoding avoids degree greater than two, as we encode squares in linear arrays of binary variables ($x'$ and $y'$), and square terms of the cost function $x'^2$, $y'^2$ and multiplication $x'y'$ are quadratic.

\bigskip\noindent
Using this approach, we have been able to factorize a 24 bits number $N=8,689,739=2958^2-245^2$ on a real quantum annealer. As far as we know, this is the biggest number ever factorized using a quantum device. Larger factorization seems tedious due to actual quantum hardware limitations.

\addcontentsline{toc}{section}{Complexity analysis}
\subsubsection*{Complexity analysis}
Let us start with the binary encoding of $x^2$. In the worst case, we have $p_{max}=2^{log_2(\sqrt{N})}-1$ and $q_{max}$ in the same order as $p_{max}$, thus $x_{max}^2=(\frac{p_{max}+q_{max}}{2})^2$, which leads to a space complexity of 
$$
\mathcal{O}{(x_{max}^2)}=\mathcal{O}{(log_2(2^{2log_2(\sqrt{N})}))}=\mathcal{O}{(2log_2(\sqrt{N}))}=\mathcal{O}{(log_2(N))}
$$

\medskip\noindent
For the binary encoding of $y_{max}^2$, in the worst case we have $q_{min}=2^{log_2(\sqrt{N})-1}+1$, thus $y_{max}^2=(\frac{p_{max}-q_{min}}{2})^2=(\frac{2^{log_2(\sqrt{N})}-1-(2^{log_2(\sqrt{N})-1}+1)}{2})^2=(\frac{\sqrt{N}-\frac{\sqrt{N}}{2}-2}{2})^2=(\frac{\sqrt{N}}{4}-1)^2=\frac{N}{16}-\frac{\sqrt{N}}{2}+1$, finally, the space complexity for $y^2$ is
$$
\mathcal{O}{(y^2)}=\mathcal{O}{(log_2(\frac{N}{16}-\frac{\sqrt{N}}{2}+1))}
$$

\medskip\noindent
The total space complexity for the encoding of $x^2$ and $y^2$ is 
$$
\mathcal{O}{(log_2(N)+log_2(\frac{N}{16}-\frac{\sqrt{N}}{2}+1))}=\mathcal{O}{(log_2(\frac{N^2}{16}-\frac{N^\frac{3}{2}}{2}+N))}=\mathcal{O}{(log_2(N))}
$$

\section{Conclusion}\label{sec3}

In this article, we present a novel improvement of the classical Fermat's method, as well as two different approaches of a quantum implementation of this method, leading to the factorization of the largest composite odd number on a real quantum hardware. As we are far from running this algorithm on a realistic RSA instance, and there is no general proof of speedup by quantum annealing over classical solvers on optimization problems, the purpose of this work is mostly to give a new perspective on the factoring problem, which can be the starting point of further study and future enhancement.

\backmatter





\bibliography{sn-bibliography}% common bib file

\end{document}